\documentclass[
  twocolumn,
  prl,
  amsmath,
  amssymb,
  superscriptaddress,
  nofootinbib,
  floatfix
]{revtex4}

\usepackage{mathtools}
\usepackage{bm}
\usepackage{dsfont}
\usepackage{graphicx}
\usepackage{pifont}
\usepackage{textcomp}
\usepackage{gensymb}
\usepackage{accents}
\usepackage{upgreek}
\usepackage{array}
\usepackage{hhline}
\usepackage{tabularx}
\usepackage{color}

\makeatletter
\def\NAT@bibsetnum#1{%
 \setlength{\topsep}{\z@}%
 \NATx@bibsetnum{#1}%
}%
\renewenvironment{thebibliography}[1]{%
 \NAT@thebibliography{#1}%
 \@clubpenalty\clubpenalty
 \let\@TBN@opr\present@bibnote
 \@FMN@list
}{%
 \@endnotesinbib
 \edef\@currentlabel{\arabic{NAT@ctr}}%
 \NAT@endthebibliography
 \global\let\auto@bib\@empty
}
\newcommand*{\supplementarystart}{%
  \close@column@grid%
  \clearpage%
  \onecolumngrid%
  \setcounter{enumiv}{0} 
  \setcounter{equation}{0} 
  \setcounter{figure}{0} 
  \setcounter{table}{0} 
  \setcounter{page}{1}
  \c@secnumdepth=4
  \renewcommand{\theequation}{s\arabic{equation}} 
  \renewcommand{\bibnumfmt}[1]{[s##1]} 
  \renewcommand{\@onlinecite}{s\citealp} 
  \renewcommand{\cite}[1]{{[}\onlinecite{##1}{]}}
  \renewcommand{\thefigure}{s\arabic{figure}}
  \renewcommand{\thetable}{s\Roman{table}}
  \renewcommand{\thepage}{s\arabic{page}}
}
\makeatother

\newcommand{\be}{\begin{equation}}
\newcommand{\e}{\end{equation}}
\newcommand{\beml}{\begin{subequations}}
\newcommand{\eml}{\end{subequations}}
\newcommand{\beq}{\begin{eqnarray}}
\newcommand{\eq}{\end{eqnarray}}
\newcommand{\ba}{\begin{array}}
\newcommand{\ea}{\end{array}}
\newcommand{\bpm}{\begin{pmatrix}}
\newcommand{\epm}{\end{pmatrix}}
\newcommand{\bc}{\begin{cases}}
\newcommand{\ec}{\end{cases}}

\renewcommand{\log}{\mathop{\mathrm{ln}}\nolimits}

\begin{document}
\title{Anomalous Hydrodynamics in One Dimensional Electronic Fluid }
\date{\today}
\author{I.~V. Protopopov}
\affiliation{Department of Theoretical Physics, University of Geneva, 1211 Geneva, Switzerland  }
\affiliation{Landau Institute for Theoretical Physics, 119334 Moscow, Russia}
\author{R. Samanta}
 \affiliation{Department of Physics, Bar Ilan University, Ramat Gan 52900, Israel}
 \author{A.~D. Mirlin}
\affiliation{Institute for Quantum Materials and Technologies, Karlsruhe Institute of Technology, 76021 Karlsruhe, Germany}
\affiliation{Institut f\"{u}r Theorie der Kondensierten Materie, Karlsruhe Institute of Technology, 76049 Karlsruhe, Germany}
\affiliation{Petersburg Nuclear Physics Institute, 188350 St. Petersburg, Russia}
\affiliation{Landau Institute for Theoretical Physics, 119334 Moscow, Russia}
\author{D.B. Gutman}
 \affiliation{Department of Physics, Bar Ilan University, Ramat Gan 52900, Israel}
 
\begin{abstract}
We construct multi-mode viscous hydrodynamics for one dimensional  
spinless electrons.
Depending on the scale, the fluid has  six (shortest lengths), four (intermediate, exponentially broad regime), or three (asymptotically long scales) hydrodynamic modes. 
Interaction between hydrodynamic modes leads to anomalous scaling of physical observables and waves propagating in the fluid. In four-mode regime, all modes are ballistic and acquire KPZ-like broadening with asymmetric power-law tails. ``Heads'' and ``tails'' of the waves contribute equally to thermal conductivity, leading to $\omega^{-1/3}$ scaling of its real part. 
In three-mode regime, the system is in the universality class of classical viscous fluid
\cite{nrs,Spohn2014}. Self-interaction of the sound modes results in KPZ-like shape, while the interaction with the heat mode results in asymmetric tails.
The heat mode is governed by Levy flight distribution, whose power-law tails give rise to $\omega^{-1/3}$ scaling of heat conductivity. 
\end{abstract}
\maketitle
Understanding properties of  
 interacting electronic systems is fundamentally important across various branches of physics. 
The problem is extremely non-trivial and multifaceted due to the impact of quantum coherence and strong interactions as well as other important ingredients, including the underlying crystal lattice and/or disorder. 
Progress has been achieved by constructing effective theories for long-living modes of electronic systems. 
Such theories are universal, i.e., insensitive to microscopic details and mostly determined by
qualitative aspects such as dimensionality, symmetries, and topology.
Paradigmatic examples of effective descriptions are Landau's Fermi-liquid theory\cite{Landau}, the theory of superfluid liquids\cite{Landau1} and the theory of diffusive modes in disordered conductors\cite{AndersonFifty}.

Recent advances in experimental techniques have made available several systems\cite{Molenkamp1994,Jong1995,Bandurin2016,Moll2016,Gooth2017,Bandurin2018} realizing (in a certain temperature range) the hydrodynamic regime of electron transport. In this regime, the dynamics is dominated by electron-electron collisions (rather than by impurity or electron-phonon scattering) and can be described by a set of equations of hydrodynamic type governing the evolution of conserved densities (charge, momentum, energy, etc.). Two aspects make such systems spectacular. First, they exhibit electron transport that is profoundly different from that observed in conventional Drude conductors. It is manifested in Gurzhi effect\cite{Gurzhi1968}, spatial non-locality\cite{Levitov2016} and unconventional magnetoresistance\cite{Andreev2011,Alekseev2016}, see Refs.\cite{Narozhnyi2017,Lucas2018} for a recent review. Second, topologically-induced qualitative diversity of underlying electronic spectra 
gives rise to unconventional hydrodynamic regimes, such as relativistic hydrodynamics in graphene \cite{Hartnoll2007}.  

The hydrodynamics of one-dimensional (1D) interacting electrons is of special interest. It often involves an extended (in integrable systems even infinite) number of conserved hydrodynamic charges\cite{MatveevAndreev2018,MatveevAndreev2019,Bertini2016,Bulchandani2018,Doyon2020,Weiner2020}. Furthermore, the reduced dimensionality of the system greatly promotes hydrodynamic fluctuations\cite{andreev80}, which can invalidate the mean-field hydrodynamic description at sufficiently long scales and drive the system into a fluctuation-dominated regime characterized by non-trivial scaling of physical observables\cite{KPZ, nrs, Spohn2014, Spohn_review2015}. The relevance of fluctuational hydrodynamics and in particular of the celebrated Kardar-Parisi-Zhang model in the context of 1D electronic fluids was discussed recently in Refs. \cite{bovogangardt,RPMG}.

In this paper, we explore the full multimode fluctuational hydrodynamics of 1D spinless fermions with short-range interaction. Our focus is on real-time dynamics and on thermal transport that was probed recently in several closely related experimental setups \cite{Schwab,Meschke,Jezouin,Cui,Pierre,Altimiras2012,Yacoby2012,Grivnin2014, Cohen2019}. We confirm the frequency scaling of the thermal conductivity, $\kappa\propto \omega^{-1/3}$, advocated recently based on a self-consistent kinetic theory
of bosonic excitations (see Ref. \cite{RPMG} and references therein). We find, however, that the earlier kinetic treatment fails to predict the correct dependence of the prefactor in this scaling on temperature and other parameters of the system.

To construct the non-linear hydrodynamic description of the system, we employ the bosonization technique\cite{gogolin,stone,delft98,giamarchi} taking into account the curvature of the electronic spectrum (i.e., finite fermion mass $m$) \cite{schick68, Sakita, Jevicki_Sakita,hald}.
Relaxation processes in such a ``non-linear Luttinger liquid'' were analyzed in several works, see Ref. \cite{imambekov11} for a review. 
It was shown in Ref. \cite{bfduality} (see also Refs. \cite{Rozhkov,imambekov2009,pgom2014,Khodas2007,Lunde2007}) that at sufficiently low temperatures [$T<T_{\rm FB}\sim 1/m l^2<\epsilon_F$, where $l$ is the range of the electron-electron interaction and $\epsilon_F$ the Fermi energy], thermal excitations in a non-linear Luttinger liquid are ``composite'' fermions with renormalized Fermi velocity $u\sim v_F$, an effective mass $m_*\sim m$ and weak interactions vanishing in the zero-momentum limit\cite{footnote1}. The composite fermions are characterized by a long lifetime $\tau_F$,
\begin{equation}
\tau_F^{-1}\sim l^4 T^7 / m_*^2 u^8.
\label{mainTau}
\end{equation}

Focusing on this low-temperature regime, we describe the dynamics of the system by the kinetic equation for fermionic quasiparticles  
\begin{equation}
\frac{\partial N_{F}(p) }{\partial t} +v^{F}_p \frac{\partial N_{F}(p)}{\partial x}= \hat{I}_{p} [N_F]\,.
\label{kinEq}
\end{equation}
Here $N_F$ is a distribution function, $v_p^F$ is the momentum-dependent velocity of fermionic quasiparticles, and $\hat{I}$ is the collision integral. 
The hydrodynamic equations arise after the projection of the kinetic equation on the zero modes of the collision integral and are valid at scales larger than fermionic mean free path $u\tau_F$. 
The collision integral in (\ref{kinEq}) is nullified by Fermi-Dirac function $n_{F}\left(\frac{\epsilon_p-vp-\mu}{T}\right)$ with chemical potential $\mu$, temperature $T$ and the boost velocity $v$.
These three parameters of the equilibrium distribution correspond to the three exactly conserved densities of the model: particle number, energy, and momentum.  
Peculiarities of the 1D kinematics give rise however to other quasi-conserved quantities (soft modes of the collision integral). First, equilibration of the particle number between the left and right movers requires processes involving a deep hole near the bottom of the band. In the bosonic description of the Luttinger liquid, such processes correspond to the Umklapp scattering and manifest themselves only at exponentially long length scale\cite{matprb12,mick2010,Mat2012}  
\begin{equation}
L_{\rm U}\sim uT^{-3/2} \epsilon_F^{1/2}e^{\frac{\epsilon_F}{T}}.
\end{equation}
 Thus, at scales shorter than $L_{\rm U}$ the system possesses four conserved quantities (total energy, total momentum, and two chiral number densities).
Second, a detailed analysis of collision processes leading to Eq. (\ref{mainTau}) shows that in such a collision the energy and momentum exchange between the chiral sectors is 
parametrically suppressed (compared to the thermal energy or momentum) 
by a factor $(T/\epsilon_F)^2 \ll 1$ \cite{supplemental}. Correspondingly, at scales shorter than 
\begin{equation}
L_4\sim\left(\frac{\epsilon_F}{T}\right)^2\frac{m_*^2u^9}{l^4T^7}
\end{equation}
the chiral sectors are effectively decoupled and six hydrodynamic modes exist in the system.

In the six-mode regime the particle densities, momentum, and energies of each chiral sector are separately conserved and we combine them into two chiral vectors
${\bf q}_\eta^T=(\rho_\eta,\pi_\eta,\epsilon_\eta)$, $\eta=R,L$. We denote by ${\boldsymbol \phi}^T_\eta=T^{-1}_\eta (\mu_\eta,v_\eta,-1)$ the vector of the corresponding conjugate thermodynamic variables. 
The conserved quantities obey the continuity equations 
\begin{equation}
\label{6modes}
\partial_t q_\eta^i+\partial_x J_\eta^i=0 \,,
\end{equation}
with index $i$ specifying the conserved charge and the corresponding flux,  
${\bf J}_\eta=(J^{\rho}_\eta,J^{\pi}_\eta,J^{\epsilon}_\eta)$.

On the linear level, one relates  
 \begin{equation}
{\bf q}_\eta(\omega,k)=\chi_\eta^{\rm ret} (\omega,k) {\boldsymbol \phi}_\eta(\omega,k)\,, 
\label{qphi}
\end{equation}
via the polarisation operator
\nolinebreak{$\chi^{\rm ret}_{i,j;\eta}(x,t)=-i\theta(t)\langle [\hat{{\bf q}}_{i,\eta}(x,t), \hat{{\bf q}}_{j,\eta}(0,0)] \rangle$}. 
Similarly, currents can be represented in terms of current response function $M$, 
\begin{equation}
{\bf J}_\eta(\omega,k)=M_\eta(\omega, k){\boldsymbol \phi}_\eta(\omega,k)/ik \,.
\end{equation}
In the $\omega=0$, small-$k$ limit, the matrix $M_\eta(k)=(ikA +k^2 D)\chi_{\eta}$ is build out of matrices of velocities ($A$), diffusion coefficients ($D$), 
and static susceptibilities $\chi_\eta\equiv \chi^{\rm ret}_\eta(\omega=0,k\rightarrow 0)$. 
The velocity matrix $A$ and the matrix of static susceptibilities are thermodynamic quantities and can be computed straightforwardly in the approximation neglecting the composite-fermion interaction. The matrix of diffusion coefficients $D$ requires more work; it can be obtained from the linearized kinetic equation (\ref{kinEq}).
See Supplemental Material (SM) \cite{supplemental} for explicit expressions for $\chi$ and $M$. 

To incorporate non-linear effects into the hydrodynamic description, we extend the expressions for hydrodynamic currents by terms of second order in the conserved densities:
\begin{equation}
\label{hydrodynamics_Nmodes}
{\bf J}_\eta=(M_\eta/ik) \chi^{-1}_\eta {\bf q}_\eta+ \frac{1}{2}\sum_{i,j}{\bf H}_{\eta;i,j}{q}_\eta^i{q}_\eta^j\,.
\end{equation}
Here, we have take the static limit $\omega=0$ and the (vector-valued) coefficients ${\bf H}_{\eta;i,j}$ can be computed neglecting the interaction of composite fermions\cite{supplemental}.

Equations (\ref{6modes}), (\ref{qphi}), and (\ref{hydrodynamics_Nmodes}) describe the six-mode hydrodynamics that exist at short length scales, $L<L_4$. 
At longer scales, the collisions equilibrate the temperatures and the boost velocities in the two chiral sectors. 
The hydrodynamic theory of the four-mode regime can be obtained through the reduction of the six-modes equations by setting $T_L=T_R=T $, $v_L=v_R=v$ and working with 
the total energy and momentum densities, $\epsilon=\epsilon_R+\epsilon_L$ and $\pi =\pi_R+\pi_L$.

At still larger length scales, $L>L_{\rm U}$, the system reaches equilibrium with respect to particle exchange between the chiral sectors. 
The corresponding three-modes hydrodynamics can be obtained through the reduction of the four-mode theory by setting $\mu_L=\mu_R=\mu$. 

In the linear hydrodynamic approximation, the continuity equations dictate that 
\begin{eqnarray}
\label{Green_function}
\chi^{\rm ret}_\eta(\omega,k)=M_\eta \left(i\omega\chi_\eta-M_\eta \right)^{-1}\chi_\eta\,.
\end{eqnarray}
The information encoded in the polarization operator enables one to compute the full set of kinetic coefficients, accessible via linear-response measurements. 
At first glance, the non-linear terms in hydrodynamic equations are irrelevant for the discussion of such linear-response quantities. This conclusion is, however, invalidated by hydrodynamic fluctuations that were so far neglected. 
Once the fluctuations are taken into account, the nonlinear hydrodynamic couplings induce strong renormalizations of bare kinetic coefficients, totally modifying the linear-response characteristics of the system.

 To account for fluctuations we promote the hydrodynamic equations (\ref{6modes}) to the Keldysh action (of Martin-Siggia-Rose type)\cite{supplemental}.
 Since at hydrodynamic scales the system is locally at equilibrium, the fluctuation-dissipation theorem holds. Therefore, the retarded part of the polarization operator $\chi^{\rm ret}$ determines also the  
Keldysh components and thus the entire action at the Gaussian level. 
The quadratic terms in the hydrodynamic currents (\ref{hydrodynamics_Nmodes}) correspond to cubic vertices in the action.

Following Ref. \cite{Spohn_review2015}, we analyze the resulting Keldysh action of fluctuational hydrodynamics within the mode-coupling approximation\cite{footnote2}.
To perform the calculation it is convenient to pass to the eigenmodes of the linearized hydrodynamic theory.
We define a new basis ${\bf \Psi}=R {\bf q}$, where $R$ diagonalizes the velocity matrix $A$, 
 $R A R^{-1}={\rm diag}(v_1,\dots,v_N) $.
 Because of the mode separation caused by different mode velocities, only diagonal correlations survive in the long-time limit, and the Keldysh pair-correlation functions of the eigenmodes 
 \begin{eqnarray}
 f_{j}(x,t)=\langle \Psi_j(x,t) \Psi_j(0,0)\rangle
 \end{eqnarray}
 satisfy the  self-consistent Dyson equations\cite{supplemental}
\begin{eqnarray}&&
\left(\partial_t+v_j\partial_x -\tilde{D}_j\partial_x^2\right)f_j(x,t)=\int_{-\infty}^{\infty}dy\int_0^tds \nonumber \\&& 
\times f_j(x-y,t-s)\partial_{y}^2R_j(y,s)\,.
\label{Npeaks}
\end{eqnarray}
Here
\begin{equation}
R_j(y,s)=\frac{1}{T^5}\sum_{l,m=1}^N \lambda_{jlm}^2f_l(y,s)f_m(y,s) \,,
\end{equation}
$\tilde{D}_j$ are diagonal elements of the effective diffusion matrix $\tilde{D}$ describing broadening of eigenmodes, and coupling constants 
$\lambda_{jlm}$ account for the mode interaction. These constants are computed from microscopis 
parameters of the original fermionic model\cite{supplemental}.

We now employ this theory to study pulses propagation in an electronic fluid as well as its linear-response properties.

We consider the time evolution of a generic disturbance created in a limited region of the fluid. 
Due to energy relaxation for times longer than fermionic energy relaxation time $\tau_F$, any disturbance is fully projected onto eigenmodes of the collision integral. At times shorter than $L_4/u$, this yields
six hydrodynamic modes $\Psi_j$. The degree to which the modes are excited depends on the overlap of the disturbance with $\Psi_j$.
These modes give rise to six ballistic pulses propagating through the fluid.
 Due to differences between the mode velocities, $\Delta u_{ij}\equiv u_i-u_j $, the separation between the peaks growths linearly with time, $L_{ij} =\Delta u_{ij}t$. The width of each peak is broadened, within the linear hydrodynamics, by the corresponding diffusion process as $(\tilde D_{j}t)^{1/2}$. The non-linear couplings further broaden the shape of the pulses and modify their shape. 
Comparing the linear and non-linear terms, one can show that non-linear broadening dominates over the normal diffusion at scales beyond $L_{*}= L_4 (T_{\rm BF} /T)^2 \gg L_4$. 
Therefore, when fluid enters the four-mode regime, it is still governed by essentially linear theory, with conventional diffusive scaling. 

\begin{figure}[!tbp]
 \centering
 \begin{minipage}[b]{0.23\textwidth}
  \includegraphics[width=\textwidth]{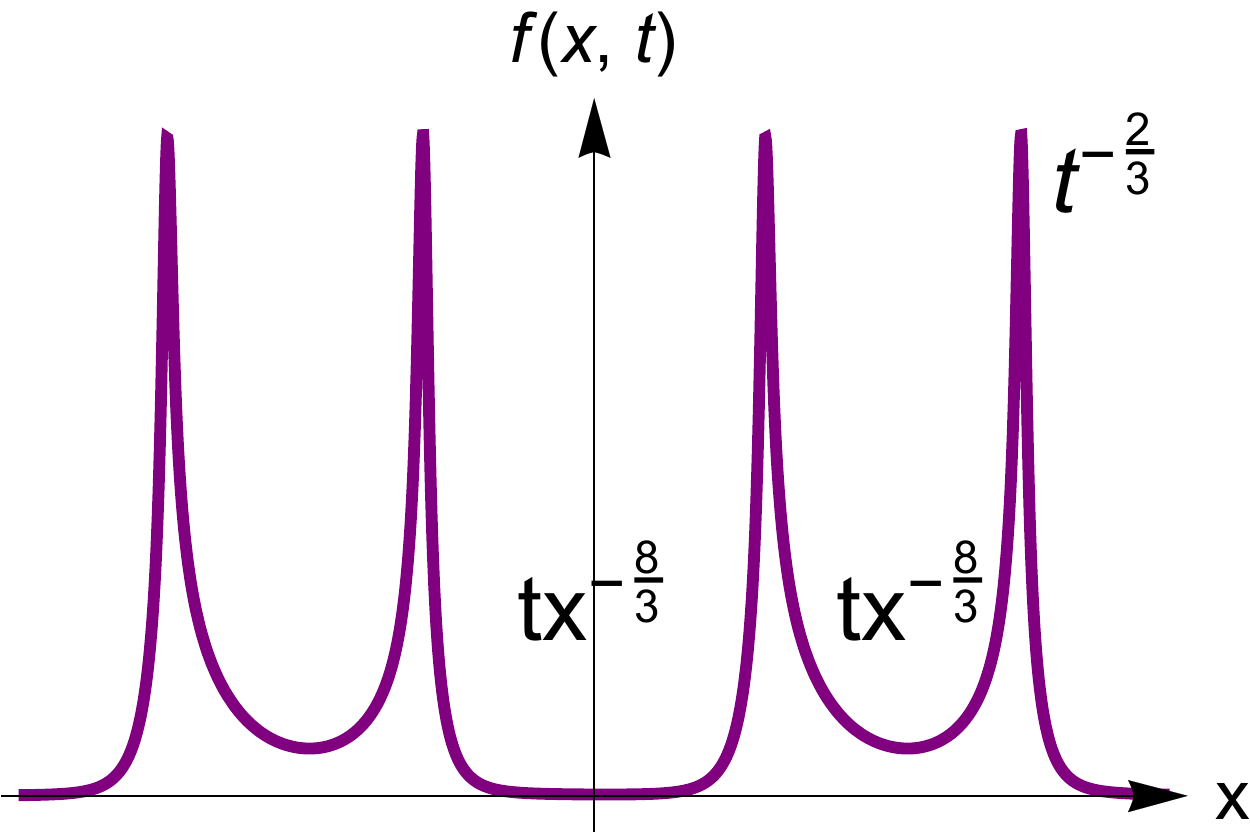}   
 \end{minipage}
 \hfill
 \begin{minipage}[b]{0.23\textwidth}
  \includegraphics[width=\textwidth]{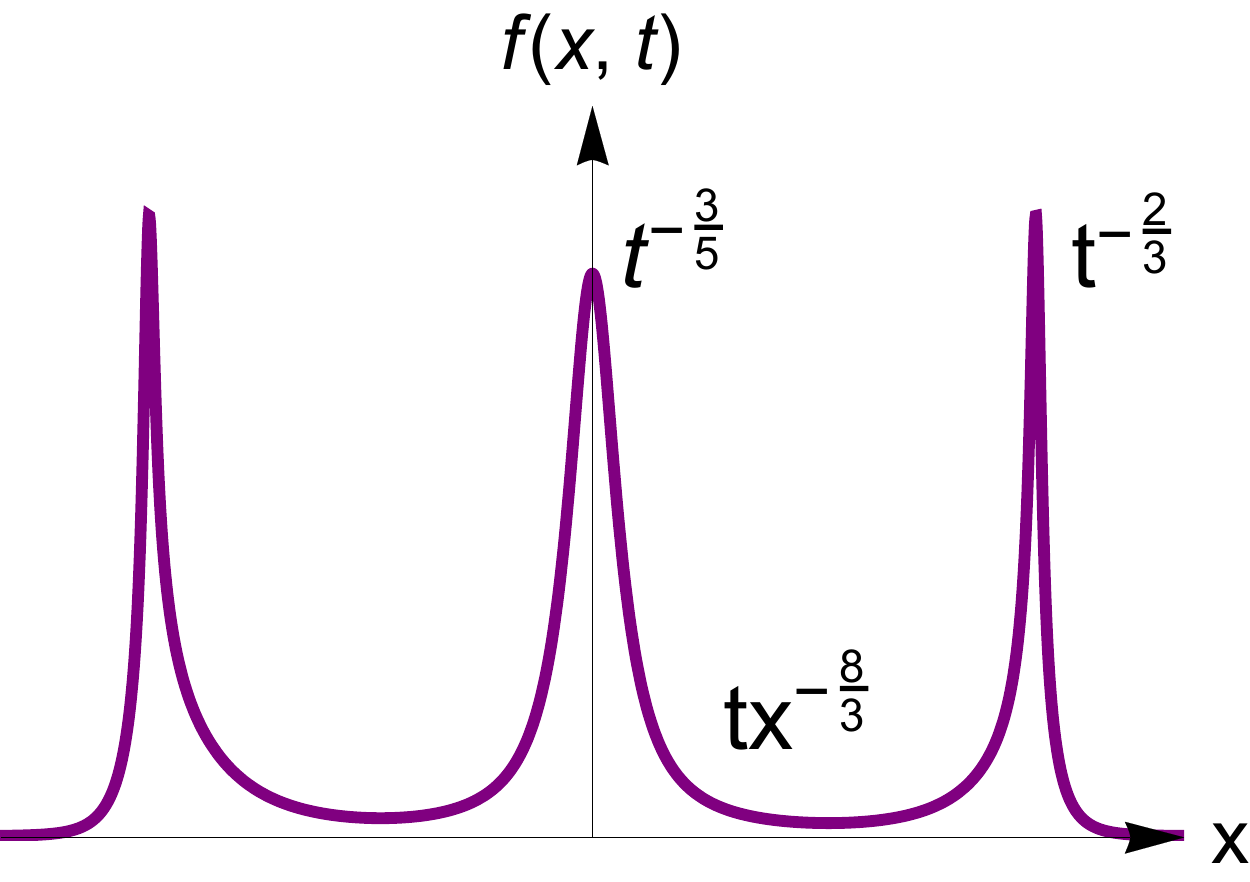}
 \end{minipage}
 \caption{Schematic shape of pulse evolution through four and three mode regimes. 
The scaling of the heads and tails of the peaks with time is depicted,  see text for details.}
\label{fig1}
\end{figure}

At $L \sim L_4$, the number of hydrodynamic modes is reduced to four and the pulses are reshaped into four peaks. The evolution in earlier stages of the four-mode regime is well described by linearized hydrodynamic. But for $L>L_*$ the non-linear terms start to dominate and the normal diffusion process is replaced by the anomalous one. 
Essentially, at this stage, one can drop the bare diffusion terms in Eq. (\ref{Npeaks}).
In the four-mode regime all non-linear coupling constants are of the same order, 
$\lambda_{ijk} \sim \lambda \equiv T u^{3/2}$. 
However, only the interaction between modes propagating in the same direction is significant.
 Therefore, Eq.~(\ref{Npeaks}) splits into two sets of chiral equations. 
Near the maximum of any given mode, the coupling to other modes is exponentially small and can be neglected. Equation (\ref{Npeaks}) is mathematically equivalent to pair velocity correlation function in the stochastic Burgers equation and the corresponding KPZ problem\cite{Beijeren2012}. 
Thus, nears the maximum
\begin{equation}
\label{KPZ_peak}
f_i(x,t)\sim \frac{T^2}{(\lambda t)^{2/3}} f_{\rm KPZ}
\left(
\frac{T(x-u_it)}{(\lambda t)^{2/3}}
\right).
\end{equation}
Here $f_{\rm KPZ}(x)$ is the universal dimensionless KPZ function, with $f_{\rm KPZ}(x) \sim 1$ for $|x| \leq 1$ and 
$f_{\rm KPZ}(x) \sim e^{-0.3 |x|^3}$ for $|x| \gg 1$\cite{Prahofer,Prahofer_Spohn}.
Away from the maximum
 the interaction between mode plays a role and creates non-symmetric power-law tails\cite{supplemental}, see Fig.\ref{fig1}.
 The fast modes develop power-law rear tails, while slow modes develop power-law front and rear tails:
\begin{eqnarray}
f_i(x,t) &\sim& \sum_j \theta[(x-u_it){\rm sgn}(\Delta u_{ji})] u^2 \left(\frac{T}{|\Delta u_{ji}|}\right)^{1/3}\nonumber 
\\&\times& t|x-u_it|^{-8/3} \ \ {\rm for} \,\, |x-u_it| \gg \frac{u t^{2/3}}{T^{1/3}}.
\label{tails}
\end{eqnarray}
One may interpret this as propagation of one degree of freedom away from its light cone via the interaction with a faster or slower degree of freedom.
In Eqs.~\eqref{KPZ_peak}, \eqref{tails} and below we omit numerical coefficients of order unity, as emphasized by the sign $\sim$ replacing the equality sign.

 At distances larger than $L_U$, the fluid is described by three hydrodynamic modes.
 This is a universal regime representing the ultimate infrared fixed point of any non-integrable system.
 It is characterized by two ballistic sound modes (index $j=2,3$) and one static (i.e., zero-velocity) heat mode ($j=1$). 
The pulse propagation in such a regime was analyzed in the context of classical fluids in Refs.~\cite{Spohn2014,Spohn_review2015,PhysRevLett.98.184301}.
The sound mode acquires the KPZ shape, Eq.(\ref{KPZ_peak}). For the corresponding self-coupling constant we find
$\lambda \equiv \lambda_{222} \sim T^4/m^3 u^{9/2}$.
 Due to the time-reversal symmetry, the self-coupling of the heat mode is identically zero. 
Therefore, in the absence of the inter-mode coupling, the spread of the heat mode would be diffusive.
The non-linear interaction between the heat and sound mode, which is characterized by a coupling 
$\lambda_{122} \sim T^3 / m^2u^{5/2}$,  
 leads to the formation of power-law tails for the heat and sound modes\cite{supplemental}.
It transforms the heat mode into symmetric Levy-flight distribution with $\alpha=5/3$,
\begin{equation}
f_1(x,t)\sim \frac{T}{\delta x(t)}
f_{{\rm Levy},\alpha=5/3}\left(\frac{x}{\delta x(t)}\right)\,.
\end{equation}
The heat mode has a maximum at $x=0$ and the width
$\delta x(t) \sim t^{3/5} T^{4/5}m^{-6/5}u^{-7/5}$.
The $t^{3/5}$ scaling of the width was also obtained in the 
context of classical anharmonic chains\cite{prvz,lukspohn}. 
The value at the maximum is $f_1(0,t)\sim T/\delta x$.
 Away from the maximum (for $x \gg \delta x$)  
 the heat mode has power law tails\cite{Zolotarev} that scale as 
$f_1(x,t)=\frac{T^{7/3}}{m^2u^{7/3}} tx^{-8/3}$, implying anomalous heat diffusion.
  
We now consider the linear response properties of the electronic fluid. 
Generally speaking, an $N$-component liquid has $N(N-1)/2$ independent linear response coefficients, that can be computed via Kubo formula\cite{supplemental}. 
The anomalous scaling observed in the pulse propagation problem manifests itself through the linear response coefficients as well.
To be specific, we focus on thermal conductivity, a quantity that describes the rate of irreversible heat propagation.
To compute the thermal conductivity, one needs first to define the heat current.
In interacting many-body problems, expressions for heat currents are in general rather complicated and spatially non-local. 
Luckily, the operator of heat current $J_T$ for fluids is local and can be computed   
by subtracting an advective contribution from the energy current $J_E$,
\cite{Luttinger1964} 
\begin{eqnarray}
J_T=J_E-\bar{w}J_\rho \,,
\end{eqnarray}
where $\bar{w}$ is the enthalpy of the fluid per one electron and $J_\rho$ the particle current.
The Kubo formula for thermal conductivity reads\cite{Castellani1987}
\begin{eqnarray}
\label{Kubo}
\sigma_{T}(\omega,k)=
\frac{1}{-i\omega T} 
\big[K_{\rm TT}(\omega,k)-K_{\rm TT}(0,0)\big],
\end{eqnarray}
where $K_{\rm TT}(\omega,k)=-i\langle [\hat{J}_T(x,t), \hat{J}_T(0,0) ]\rangle^{\rm ret}(\omega,k)$.
Employing Eq.(\ref{Kubo}) and setting $k=0$, we find \cite{supplemental} that in the 6-mode regime
\begin{equation}
\sigma_T(\omega)= - \frac{\pi^2 uT}{3i\omega}+\frac{\pi^2 u}{6m^2l^4 T^2}\,.
\end{equation}
The Drude peak corresponds to the ballistic propagation of heat\cite{kane96,lev2010}, 
while the real part of conductivity is due to heat diffusion.
As the system enters the four-mode regime, the propagation of all modes remains ballistic. Hence, the imaginary part of the heat conductivity is unchanged, 
${\rm Im} \, \sigma_T(\omega) = \pi^2 uT / 3\omega$.
The real part of the heat conductivity, on the other hand, is renormalized. The effects of the renormalization are associated with an anomalous broadening of the pulses, with
two contributions coming from the head and the tails of the peak.
Both happen to be of the same order and lead to  
\begin{equation}
{\rm Re} \, \sigma_T(\omega) \sim uT^{1/3}\omega^{-1/3}\,.
\label{thermal_4}
\end{equation}
Finally, we discuss the three-mode regime,
where the heat conductivity is determined solely by the static mode.
Therefore, the ballistic contribution is suppressed, giving rise to an exponentially large constant: $i /\omega \mapsto \tau_U$. 
The real part of the thermal conductivity thus scales as 
 \begin{equation}
 \label{thermal_3}
 {\rm Re} \ \sigma_T(\omega) \sim uT\tau_U+ \frac{T^{7/3}}{m^2 u^3}\omega^{-1/3}.
\end{equation}

To summarize, we have developed a multi-mode hydrodynamic approach for the electronic fluid.
Depending on the number of conserved charges, the fluid has six, four, or three hydrodynamic modes.
Though the three-mode regime is an ultimate long-distance fixed point, it is only reached at exponentially long distances, leaving room to
an exponentially long four-mode viscous hydrodynamic regime.

The interaction between the hydrodynamic modes leads to the renormalization of transport coefficients, giving rise to universal scaling behavior, and 
shapes pulses propagating through the fluid.  
In the six- and four-mode regimes, all pulses propagate ballistically.  
The ``head'' of every pulse is controlled by self-interaction, resulting in a KPZ scaling of pulse width ($t^{2/3}$) and amplitude ($t^{-2/3}$) with time.
 Interaction between the modes propagating with different velocities results in power-law tails scaling as $x^{-8/3}$ with distance $x$ from the mode center and directed towards another mode. 
As the system reaches a three-mode regime, the pulses redistribute, and a static heat and two ballistic sound peaks are formed.
The width of the ballistic modes has KPZ scaling with time. The interaction between the sound waves and the heat mode gives rise to power-law tails for all peaks. Each sound mode acquires a rear tail. The static heat mode is a Levy flight function with $\alpha=5/3$, with symmetric tails.
The anomaly in peak shapes leads to anomalous kinetic coefficients, in particular, the thermal conductivity. 

We conclude by comparing the results of the present analysis for $\sigma_T$ with earlier calculations performed within the self-consistent 
kinetic approach\cite{RPMG}. Reassuringly, both approaches yield two regimes of anomalous scaling of $\sigma_T$ separated by the scale $L_U$.
Further, the kinetic approach yields for these regimes results analogous to Eqs.~\eqref{thermal_3} and \eqref{thermal_4}, with the same $\omega^{-1/3}$ scaling. 
Such agreement in scaling resulting from self-consistent kinetic\cite{andreev80,Samokhin,bovogangardt} and classical renormalization-group \cite{nrs} approaches has been known for a long time. 
This agreement is highly non-trivial and perhaps even puzzling. Indeed, although our starting point here is a transport coefficient computed within fermionic kinetic theory\cite{RPMG,Matveev2019}, the subsequent analysis in this work and Ref.~\cite{RPMG} is very different. In the kinetic framework of Ref.~\cite{RPMG}, the $\omega^{-1/3}$ scaling results from subthermal bosons with wave vectors $k \ll T/u$ that can propagate anomalously large distances without scattering. At the same time, in the present framework, this enhancement of thermal conductivity results from the interaction between bosonic (hydrodynamic) modes leading to anomalous hydrodynamics. Importantly, while the frequency scaling agrees in two approaches, the prefactors (in particular, the temperature scaling) are essentially different. 
Effects of renormalization controlling results of the present work turn out to be dominant for $\sigma_T$, in both four-mode and three-mode regimes.

Note added: While preparing the paper for publication, we learnt about a recent paper \cite{Matveev2020}
that has some overlap with the present work.


\textit{ Acknowledgements} 
 D. G. was supported
by ISF-China 3119/19 and ISF 1355/20.

\supplementarystart
\centerline{\bfseries\large ONLINE SUPPLEMENTAL MATERIAL}
\vspace{6pt}
\centerline{\bfseries \large Anomalous Hydrodynamics in One Dimensional Electronic Fluid}
\vspace{6pt}
\vspace{6pt}
\centerline{I.V.  Protopopov, R. Samanta, A.D. Mirlin and D.B. Gutman}

\section{Boltzmann equation and kinetic coefficients}
In this section of SM we derive the hydrodynamic description of the fluid in the 6-mode regime from 
the kinetic  theory, establish the relation between currents and the zero modes of the collision integral, needed for  Eq.(6)in the  main text;  express currents via  thermodynamic variables, see Eq.(8) in the main text and  evaluate  the bare diffusion coefficients.

Our starting point is a  kinetic equation formulated in terms of "composite" 
electrons\cite{Rozhkov,Samanta,PGOM}.
\begin{eqnarray}
\frac{\partial f(p,x,t)}{\partial t}+v_p\frac{\partial f(p,x,t)}{\partial x} =\hat{I}[f]\,.
\end{eqnarray}
We now outline this approach to compute linear response. Close to local equilibrium the distribution function can be written as  
\begin{equation}
\label{dist_function}
f(p,x,t)=n_{F}\left(\frac{\epsilon_p-v(x,t)p-\mu(x,t)}{T(x,t)}\right)+\delta f(p,x,t)\,,
\end{equation}
where $n_F$ is the Fermi-Dirac distribution.
Here the first term corresponds to the zero mode of the collision integral and the second to  ``massive'' excitations. On scales longer than the inelastic collision length, the two parts are related via 
\begin{eqnarray}
\delta f(p,x,t)=\frac{1}{4T}\int (dp')\hat{I}_{p,p'}^{-1}v_{p'}g^2(p')\left(\partial_x\mu-\frac{\mu}{T} \partial_xT+p'\partial_xv\right).
\end{eqnarray}
Here we define $g(p)=\frac{1}{2}\cosh^{-1}\left(\frac{\epsilon_p-\mu}{2T}\right)$, and $\hat{I}^{-1}_{p,p'}$ is an operator that is inverse to a
linearised collision integral.  
The $\hat{I}_{p,p'}^{-1}$ is not symmetric with respect to $p\leftrightarrow p'$. 
However by a transformation  
\begin{equation}
\hat{I}^{-1}_{p,p'} =g(p)\hat{{\cal I}}^{-1}_{p,p'}g^{-1}(p')
\end{equation}
it is related to a symmetric operator $\hat{{\cal I}}^{-1}_{p,p'}$.  
We now use the eigenfunctions of $\hat{{\cal I}}^{-1}_{p,p'}$ to classify the spectrum of kinetic equation, 
and separate the currents into ideal and dissipative parts 
\begin{equation}
J_\alpha(k)=J^{\rm ideal}_\alpha(k)+J^{\rm diss}_\alpha(k)\,.
\end{equation}

The dissipationless (ideal) part of the current is carried by gapless excitations
that are parametrized by time and space-dependent zero modes ($T,\mu, v$).
These excitations give rise to particle, momentum and energy currents that on the linear level are given by
\begin{equation}
J^{\rm ideal}_{\alpha}(k)=\frac{1}{4}\int (dp)\hat{j}_{\alpha}(p)g^2(p)\bigg[{\boldsymbol \phi}_1(k)+p{\boldsymbol \phi}_2(k)+
(\epsilon_p-\mu){\boldsymbol \phi}_3(k)\bigg]\,.
\end{equation}
Here $\alpha=\{\rho,\pi,\epsilon\}$ and we denote $\hat{j}_{\rho}(p)=v_p, \hat{j}_{\pi}=pv_p, \hat{j}_{\epsilon}(p)=(\epsilon_p-\mu) v_p$.
The zero modes are combined into 
\begin{eqnarray}
{\boldsymbol \phi}^T_\eta=T^{-1}_\eta (\mu_\eta,v_\eta,-1).
\end{eqnarray}

The currents above rely solely on the thermodynamic properties of the system. For this reason, they are identical to corresponding currents in the ideal fermionic hydrodynamics\cite{PGSM}, with identical conserved quantities. 
In particular, in the six-mode regime, the chiral currents agree with chiral currents of hydrodynamics on one-dimensional edge of a quantum Hall sample.  
 
In addition to the zero modes, there are finite-energy modes of the collision integral that  lead to dissipative processes.
They correspond to following currents
\begin{eqnarray}&&
J^{\rm dis}_\alpha(k)=\frac{ik}{4}\int (dp)(dp')j_\alpha(p)g(p){\cal I}^{-1}_{p,p'}g(p')v_{p'}
\bigg[{\boldsymbol \phi}_1(k)+p'{\boldsymbol \phi}_2(k)+(\epsilon_{p'}-\mu){\boldsymbol \phi}_3(k) \bigg]=\nonumber \\&& 
=ik
\big[ d_{\alpha,1}{\boldsymbol \phi}_1(k)+ d_{\alpha,2}{\boldsymbol \phi}_2(k)+ d_{\alpha,3}{\boldsymbol \phi}_3(k)\big]\,.
\end{eqnarray}
Here we define  
\begin{equation}
d_{\alpha,\beta}=-\frac{1}{4}\int (dp)(dp')\hat{j}_\alpha(p) g(p)\hat{{\cal I}}^{-1}_{p,p'}g(p')\hat{j}_\beta(p').
\end{equation}

By construction, the operator $\hat{{\cal I}}^{-1}$ is computed in the space orthogonal to zero modes.
Therefore, to compute the dissipative contribution one needs to project the current vertices $j_\alpha(p)$ onto a subspace
that is orthogonal to zero modes. In the six-mode regime, the zero modes correspond to the conservation of chiral densities, momenta, and energies.
This corresponds to functions $g(p)$, $pg(p)$ and $(\epsilon(p)-\mu)g(p)$.
We now focus on the {\it dc} limit $\omega \rightarrow 0$ and show how kinetic coefficients can be computed in this framework. 
In the six-mode regime  one finds, for the right moving electrons,
\begin{eqnarray} &&
\label{currents}
2\pi J_\rho^R(k)=
\big[T -ik d_{11}\big]{\boldsymbol \phi}_1(k)+
\big[mu T -\frac{\pi^2T^3}{6mu^3} -ik d_{12}\big]{\boldsymbol\phi}_2(k) 
+\big[-ik d_{13}\big]{\boldsymbol \phi}_3(k) \,,  \nonumber
\\&&
2 \pi J_\pi^R(k)=\big[mu T-\frac{\pi^2T^3}{6 mu^3}-ik d_{21}\big] {\boldsymbol \phi}_1(k)+
\big[m^2u^2 T -ik d_{22}\big]{\boldsymbol \phi}_2(k)+\big[\frac{\pi^2T^3}{3u}-ik d_{23}\big] {\boldsymbol \phi}_3(k)\,,  \nonumber \\&&
2\pi J_\epsilon^R(k)=\big[ -ik d_{31}\big]{\boldsymbol \phi}_1(k)+
\big[\frac{\pi^2T^3}{3u} -ik d_{32}\big] {\boldsymbol \phi}_2(k) +
\big[\frac{\pi^2T^3}{3}-ikd_{33}\big]
 {\boldsymbol \phi}_3(k) \,.
\end{eqnarray}

It is worth noting that for strictly parabolic spectrum only the thermal conductivity $d_{33} \sim \frac{u}{m^2 l^4 }$ is finite, 
while the rest of the elements $d_{\alpha,\beta}=0$. 
In the kinetic theory of nearly ideal gas, it is well known that its viscosity coefficient vanishes for particles with a strictly parabolic spectrum \cite{Lifshits_Pitaevskii}. For $d=1$ we find that the real part of all kinetic coefficients, except thermal conductivity, also vanishes.

The relation above yield current $J_\alpha$ as linear functions of the external forces
${\boldsymbol \phi}$, thus defining a matrix of linear response coefficients $\hat{L}$
\begin{eqnarray}
\label{thermal_current}
\langle J_n\rangle(\omega,k)=i k T \bigg
[L_{n,1}(k){\boldsymbol \phi}_1(\omega,k)+L_{n,2}(k){\boldsymbol \phi}_2(\omega,k)+L_{n,3}(k){\boldsymbol \phi}_3(\omega,k)\bigg].
\end{eqnarray}

It is useful to express the current operators in terms of thermodynamic variables.
Resolving the relation between the zero modes of the collision integral ${\boldsymbol \phi}$ and thermodynamic variables  
$\rho^R,\pi^R, \epsilon^R$
one extends Eq. (\ref{currents}) onto non-linear level
\begin{eqnarray} &&
J_\rho^R=\frac{1}{m}\pi^R \,, \\&&
J_\pi^R=2\epsilon^R+mu^2\rho^R \,, \\&&
J_\epsilon^R \simeq 2mu^3\rho^R-2u^2\pi^R+3u\epsilon^R -ikD_\epsilon
\left(\epsilon^R+mu^2\rho^R-u\pi^R\right) +\Delta J^R_\epsilon\,.
\end{eqnarray}
Here we have defined the energy diffusion constant $D_\epsilon=u^6/l^4T^5$.
It is worth mentioning that the results for the particle $J_\rho$ and momentum $J_\pi$ currents are exact.
There are no non-linear in density correction to these currents, neither $T/mu^2$ corrections.
The results for the energy current contains only the leading in $T/mu^2$ part.
There is also a non-linear in density-field part,  which is given by
\begin{equation}
\Delta J^R_\epsilon\simeq-\frac{99225 m^2 u^6 }{128 \pi ^3 T^4}
(\epsilon^R+mu^2\rho^R -u\pi^R)^2\,.
\end{equation}
The linear-response coefficients $L_{ij}$ computed above on the linear level, see Eq.(\ref{currents}),
are affected by non-linearities that have been so far ignored. The latter affect the character of heat propagation, or on a more technical level induce a renormalization of the theory. 
This necessitates development of a non-linear field theory of multi-component fluid.
We now promote the equations of motion to the level of action, formulating an effective classical field theory\cite{kamenev,rammer}.

\section{Martin-Siggia-Rose action}
In this section we describe the classical field theory that corresponds to the fluctuation  hydrodynamics of electron fluid, and derive mode-coupling equation, see  Eq.(11) in the main text.

We start with Gaussian part of the action 
\begin{equation}
\label{S0}
S_0=({\bf q}^T,\bar{{\bf q}}^T)_{-\omega,-k}T \hat{\chi}^{-1}_{\omega,k} \left(\begin{matrix}
 {\bf q} \\ 
 \bar{{\bf q}}
\end{matrix}\right)_{\omega,k}\,.
\end{equation}
Here we denote
\begin{equation}
\hat{\chi}^{-1}_{\omega,k}= \left(\begin{matrix}
 0 & \hat{\chi}^{{\rm a}^{-1}}_{\omega,k} \\ 
 \hat{\chi}^{{\rm r}^{-1}}_{\omega,k} & (\hat{\chi}^{{\rm r}^{-1}}_{\omega,k} -\hat{\chi}^{{\rm a}^{-1}}_{\omega,k})B_\omega
\end{matrix}\right)\,,
\end{equation}
where $B_\omega=\coth\left(\frac{\omega}{2T}\right)$.
The action is encoded by
\begin{equation}
\chi_\eta^{{\rm r}^{-1}}(\omega,k)=i\omega M_\eta^{-1}(k)+\chi^{-1}_{0,\eta}(k)\,.
\end{equation}
In the limit $\omega \rightarrow 0$ the expansion 
\begin{equation}
M_\eta(k)\chi^{-1}_{0,\eta}(k)=ikA_\eta+k^2 D_\eta
\end{equation}
yield the generalised velocity $A$ and diffusion $D$ matrices. 
Their size is equal to the number of conserved modes.  
Because the system is in local equilibrium, analytic properties and fluctuation-dissipation theorem allow to restore advance 
and Keldysh components,
$\chi^K(\omega,k)=[\chi^{\rm ret}(\omega,k)-\chi^{\rm adv}(\omega,k)]\coth\frac{\omega}{2T}$.
The interaction part of the action
\begin{equation}
S_{\rm int}=-iT \sum_{p,l,m}\bar{q}^p\Gamma_{p,l,m}q^lq^m\,,
\end{equation}
and the interaction vertex
\begin{equation}
\Gamma_{p,l,m}(k)=k\sum_{p_1}\hat{M}^{-1}_{p,p_1}(k)H^{p_1}_{l,m} (k)\,.
\end{equation}
Here 
\begin{equation}
H^{p}_{l,m}(k) =\frac{\partial^2 J_p(k)}{\partial q_l\partial q_m}\,.
\end{equation}
Because matrices $M$ and $\chi$ are symmetric, the Keldysh action (\ref{S0}) can be diagonalized as a quadratic form 
by the linear transformation 
\begin{equation}
\label{rotation_R}
{\bf q} = R^{-1} {\bf \Psi}\, \, , \bar{{\bf q}}=R^{-1} \bar{{\bf \Psi}} \,.
\end{equation}
Since the compressibility matrix has positive eigenvalues, one can choose
\begin{eqnarray}
\frac{1}{T}R \chi_0 R^T=\hat{1}
\end{eqnarray}
and 
\begin{eqnarray}
R A R^{-1}=\hat{v}\,.
\end{eqnarray}
Here $\hat{v}={\rm diag}(v_1,v_2, \dots)$ is the diagonal velocity matrix.

After the rotation (\ref{rotation_R}), the retarded part of the Gaussian action takes the form
\begin{equation}
\label{spsi}
S_0^{\rm ret}[{\bf \Psi}]=\sum_{m,n}\bar{{\bf \Psi}}^T_m(-\omega,-k)\bigg[i\omega\left(i\hat{v}k+k^2\tilde{D}\right)^{-1}+\hat{1}]_{m,n}{\bf \Psi}_n(\omega,k)\,,
\end{equation}
where $\tilde{D}=RDR^{-1}$ is the diffusuion matrix in the eigenmode basis.
The corresponding retarded propagator reads
\begin{equation}
\label{psi_retarded}
i\langle \bar{{\bf \Psi}}_m{\bf \Psi}_n\rangle_{\omega,k} =
\left(i\omega (i\hat{v} k+\tilde{D} k^2)^{-1}+1\right)^{-1}_{m,n}\,.
\end{equation}
The advanced propagator is related to the retarded one via
\begin{eqnarray}
\label{psi_advanced}
i\langle {\bf \Psi}_m\bar{{\bf \Psi}}_n\rangle_{\omega,k} =-i\langle \bar{{\bf \Psi}}_m{\bf \Psi}_n\rangle^*_{\omega,k} \,.
\end{eqnarray}
The Keldysh part of the propagator follows from FDT theorem\cite{Landau_SP1}
\begin{equation}
\label{psi_Keldysh}
\langle {\bf \Psi}_m{\bf \Psi}_n\rangle(\omega,k) =
\bigg[\langle \bar{{\bf \Psi}}_m{\bf \Psi}_n\rangle(\omega,k)- \langle {\bf \Psi}_m\bar{{\bf \Psi}}_n\rangle(\omega,k)\bigg]_{m,n}\coth\frac{\omega}{2T}\,.
\end{equation}
From here one infers equal time correlation function in x-representation 
$ \langle {\bf \Psi}_k(x,0) {\bf \Psi}_k(0,0)\rangle =T\delta(x)$. 

Next, we discuss the non-linear part of the action.
In terms of eigenmodes ${\bf \Psi}$ it reads
\begin{equation}
\label{spsiint}
S_{\rm int}=-i \sum_{k,l,m} \gamma_{k,l,m}\bar{{\bf \Psi}}^T_k {\bf \Psi}_l{\bf \Psi}_m\,,
\end{equation}
where  
\begin{equation}
\gamma_{k,l,m}= T\sum_{k_1,l_1,m_1}R_{k,k_1}^{-1^T}\Gamma_{k_1,l_1,m_1} R^{-1}_{l_1,l} R^{-1}_{m_1,m} \,.
\end{equation}
By varying the action with respect to the field $\bar{{\bf \Psi}}_k$, one derives the equation of motion
\begin{equation}
\left(\partial_t-v_k\partial_x-\tilde{D}_k\partial_x^2 \right){\bf \Psi}_k(x,t)-\frac{1}{T^2}\sum_{l,m} \lambda_{k,l,m} \partial_x\left({\bf \Psi}_l(x,t){\bf \Psi}_m(x,t)\right)=0\,.
\end{equation}
Here
\begin{equation}
\lambda_{p,l,m}=-iT^2\sum R_{p,p_1}{H^{p_1}_{l_1,m_1}} (R^{-1})_{l_1,l}(R^{-1})_{m_1,m}\,.
\end{equation}
Multiplying this equation by ${\bf \Psi}_k(0,0)$ and averaging over the action, 
one gets an equation of motion for the correlation function
\begin{equation}
\left(\partial_t-v_k\partial_x-\tilde{D}_k\partial_x^2\right)\langle {\bf \Psi}_k(x,t){\bf \Psi}_k(0,0)\rangle-\frac{1}{T^2}
\sum_{l,m}\lambda_{k,l,m}\partial_x\langle {\bf \Psi}_l(x,t){\bf \Psi}_m(x,t){\bf \Psi}_k(0,0)\rangle=0\,.
\end{equation}
Employing the Wick's theorem, one expresses the triple correlation function in terms of the pair correlation function.
Performing the self-consistent approximation, one finds the mode-coupling equation\cite{Spohn2014}. 

\section{response functions from microscopics}
In this part of SM we derive the response coefficients that appear in Eqs. (7) and (8) 
of the main text for the 6,4 and 3-mode regimes. We also find the eigen-modes  $\Psi$ of linearised hydrodynamics in all the regimes, defined in the main text, compute their velocities  ($v_j$)and diffusion matrices ($\tilde{D}$), 
and coupling constants ($\lambda_{ijk}$) needed for  Eq.(10) in the main text. 

\subsection{6-mode regime}
In this section we derive the hydrodynamic model from the kinetics (\ref{dist_function})  for the 6-modes regime.
We start with the  susceptibility matrix, that connects ${\bf q}$ and ${\boldsymbol \phi}$ 
variables on the linear level.
Multiplying the distribution function by $1,p,\epsilon_p$ and integrating over momentum, we find 
that for the right-moving electrons
\begin{equation}
\label{chi6}
\chi_6 =\frac{T}{2\pi}\left(
\begin{array}{ccc}
 \frac{49 \pi ^4 T^4}{24 m^4 u^9}+\frac{\pi ^2 T^2}{2 m^2 u^5}+\frac{1}{u} & m & -\frac{7 \pi ^4 T^4}{6 m^3 u^7}-\frac{\pi ^2 T^2}{3 m u^3} \\
 m & -\frac{7 \pi ^4 T^4}{24 m^2 u^7}+m^2 u-\frac{\pi ^2 T^2}{6 u^3} & 0 \\
 -\frac{7 \pi ^4 T^4}{6 m^3 u^7}-\frac{\pi ^2 T^2}{3 m u^3} & 0 & \frac{7 \pi ^4 T^4}{10 m^2 u^5}+\frac{\pi ^2 T^2}{3 u} \\
\end{array}
\right)\,.
 \end{equation}
Integrating the distribution function (\ref{dist_function}) with current operators $j_\alpha(p)$ over the momentum, we compute 
the matrix of currents $M_6$. 
The low-$k$ expansion of $M_6$ reads
\begin{equation}
\label{M6}
M_6\simeq 
\frac{ik T}{2\pi} 
\left(
\begin{array}{ccc}
 1 & -\frac{7 \pi ^4 T^4}{24 m^3 u^7}-\frac{\pi ^2 T^2}{6 m u^3}+m u & 0 \\
 -\frac{7 \pi ^4 T^4}{24 m^3 u^7}-\frac{\pi ^2 T^2}{6 m u^3}+m u & m^2 u^2 & \frac{7 \pi ^4 T^4}{30 m^2 u^5}+\frac{\pi ^2 T^2}{3 u} \\
 0 & \frac{7 \pi ^4 T^4}{30 m^2 u^5}+\frac{\pi ^2 T^2}{3 u} & \frac{\pi ^2 T^2}{3} \\
\end{array}
\right)\,.
\end{equation}
We next  find the eigenmodes of linearised hydrodynamic ($\Psi$), and   compute 
the rotation matrix 
\begin{eqnarray}
R_6\simeq \frac{m u^{5/2}}{T^{2}}
\left(
\begin{array}{ccc}
 \frac{2 \sqrt{\frac{15}{7}} m u^2}{\pi ^{3/2}}+\frac{79}{16} \sqrt{\frac{3}{\pi }} T & -\frac{79 \sqrt{\frac{3}{\pi }} T}{16 m u}-\frac{2 \sqrt{\frac{15}{7}} u}{\pi ^{3/2}} & \frac{95
  \sqrt{\frac{3}{\pi }} T}{16 m u^2}+\frac{2 \sqrt{\frac{15}{7}}}{\pi ^{3/2}} \\
 \frac{79}{16} \sqrt{\frac{3}{\pi }} T-\frac{2 \sqrt{\frac{15}{7}} m u^2}{\pi ^{3/2}} & \frac{2 \sqrt{\frac{15}{7}} u}{\pi ^{3/2}}-\frac{79 \sqrt{\frac{3}{\pi }} T}{16 m u} & \frac{95
  \sqrt{\frac{3}{\pi }} T}{16 m u^2}-\frac{2 \sqrt{\frac{15}{7}}}{\pi ^{3/2}} \\
 -\frac{5 \sqrt{\frac{3}{14}} m u^2}{\pi ^{3/2}} & \frac{5 \sqrt{\frac{3}{14}} u}{\pi ^{3/2}} & -\frac{5 \sqrt{\frac{3}{14}}}{\pi ^{3/2}} \\
\end{array}
\right)\,.
\end{eqnarray}
The velocities of eigenmodes in this regime are given by
\begin{eqnarray}&&
u_1\simeq u+\frac{\sqrt{\frac{7}{5}} \pi T}{m u},\\&&
u_2\simeq u-\frac{\sqrt{\frac{7}{5}} \pi T}{m u},\\&&
 u_3\simeq u+\frac{49 \pi ^2 T^2}{32 m^2 u^3}\,.
\end{eqnarray}

For the strictly parabolic spectrum the diffusion matrix reads
\begin{equation}
\label{D}
D_6 \simeq
D_\epsilon 
\left(\begin{matrix}
0 & 0 & 0 \\
0 & 0 & 0 \\
mu^2  & u & 1
 \end{matrix}\right)\,.
\end{equation}
 In terms of eigenmodes the diffusion matrix $\tilde{D}_6= R_6 D_6 R^{-1}_6$ reads
\begin{eqnarray}
\tilde{D}_6\simeq
\frac{15 }{7\pi^4}
D_\epsilon
\left(
\begin{array}{ccc}
 -2 & 2 & -\sqrt{\frac{5}{2}} \\
 2 & -2 & \sqrt{\frac{5}{2}} \\
 -\sqrt{\frac{5}{2}} & \sqrt{\frac{5}{2}} & -\frac{5}{4} \\
\end{array}
\right)\,.
\end{eqnarray}
The right- and left-moving heat modes can be expressed in terms of hydrodynamic eigenmodes as 
\begin{eqnarray}
\epsilon_R-\bar{w}\rho_R=\frac{\pi T}{\sqrt{6u}}(\Psi_1+\Psi_2)\,, \qquad \epsilon_L-\bar{w}\rho_L=\frac{\pi T}{\sqrt{6u}}(\Psi_4+\Psi_5)\,.
\end{eqnarray}

\subsection{4-mode regime}
In this section we derive the hydrodynamic model from the kinetics for the 4-modes regime.
In this case,  the set of zero modes of the collision integral is described by
\begin{eqnarray}
{\boldsymbol \phi}^T=T^{-1} (\mu_R,\mu_L,v,-1).
\end{eqnarray}
Repeating the steps analogous to the 6-mode regime, we find the susceptibility matrix
\begin{equation}
\label{chi4}
\chi_{4} =\frac{T}{2\pi}
\left(
\begin{array}{cccc}
 \frac{49 \pi ^4 T^4}{24 m^4 u^9}+\frac{\pi ^2 T^2}{2 m^2 u^5}+\frac{1}{u} & 0 & m & -\frac{7 \pi ^4 T^4}{6 m^3 u^7}-\frac{\pi ^2 T^2}{3 m u^3} \\
 0 & \frac{49 \pi ^4 T^4}{24 m^4 u^9}+\frac{\pi ^2 T^2}{2 m^2 u^5}+\frac{1}{u} & -m & -\frac{7 \pi ^4 T^4}{6 m^3 u^7}-\frac{\pi ^2 T^2}{3 m u^3} \\
 m & -m & -\frac{7 \pi ^4 T^4}{12 m^2 u^7}+2 m^2 u-\frac{\pi ^2 T^2}{3 u^3} & 0 \\
 -\frac{7 \pi ^4 T^4}{6 m^3 u^7}-\frac{\pi ^2 T^2}{3 m u^3} & -\frac{7 \pi ^4 T^4}{6 m^3 u^7}-\frac{\pi ^2 T^2}{3 m u^3} & 0 & \frac{7 \pi ^4 T^4}{5 m^2 u^5}+\frac{2 \pi ^2 T^2}{3 u} \\
\end{array}
\right)
\end{equation}
and the matrix of currents
\begin{equation}
\label{M4}
M_{4} =ik\frac{T}{2\pi}
\left(
\begin{array}{cccc}
 1 & 0 & -\frac{7 \pi ^4 T^4}{24 m^3 u^7}-\frac{\pi ^2 T^2}{6 m u^3}+m u & 0 \\
 0 & -1 & -\frac{7 \pi ^4 T^4}{24 m^3 u^7}-\frac{\pi ^2 T^2}{6 m u^3}+m u & 0 \\
 -\frac{7 \pi ^4 T^4}{24 m^3 u^7}-\frac{\pi ^2 T^2}{6 m u^3}+m u & -\frac{7 \pi ^4 T^4}{24 m^3 u^7}-\frac{\pi ^2 T^2}{6 m u^3}+m u & 0 & \frac{7 \pi ^4 T^4}{15 m^2 u^5}+\frac{2 \pi ^2
  T^2}{3 u} \\
 0 & 0 & \frac{7 \pi ^4 T^4}{15 m^2 u^5}+\frac{2 \pi ^2 T^2}{3 u} & 0 \\
\end{array}
\right)\,.
\end{equation}
The rotating matrix
\begin{eqnarray}
R_4\simeq \frac{\sqrt{3u}}{2 \pi^{3/2} T}
\left(
\begin{array}{cccc}
 m u^2 & -m u^2 & -u & 1 \\
 -m u^2 & m u^2 & u & 1 \\
 m u^2 & -m u^2 & -u & 1 \\
 -m u^2 & m u^2 & u & 1 \\
\end{array}
\right)\,.
\end{eqnarray}
The velocities in this regime split linearly with temperature, in agreement with a general argument given in Ref.~\cite{Matveev_Andreev2018}:
\begin{eqnarray}&&
u_1=-u+\frac{\pi T}{\sqrt{3} m u} \,, \\&&
u_2=u-\frac{\pi T}{\sqrt{3} m u} \,, \\&&
u_3=-u-\frac{\pi T}{\sqrt{3} m u} \,, \\&&
u_4=u+\frac{\pi T}{\sqrt{3} m u}\,.
\end{eqnarray}
We next compute the diffusion matrix 
\begin{eqnarray}
D_4\simeq D_\epsilon\left(
\begin{array}{cccc}
 0 & 0 & 0 & 0 \\
 0 & 0 & 0 & 0 \\
 0 & 0 & 0 & 0 \\
 m u^2 & m u^2 & 0 & 1 \\
\end{array}
\right)\,.
\end{eqnarray}
After the rotation into eigen-mode basis 
$\tilde{D_4}=R_4D_4R^{-1}_4$, one finds
\begin{eqnarray}
\tilde{D}_4\simeq \frac{3 }{8\pi^2 } D_\epsilon
\left(
\begin{array}{cccc}
 1 & 1 & 1 & 1 \\
 1 & 1 & 1 & 1 \\
 1 & 1 & 1 & 1 \\
 1 & 1 & 1 & 1 \\
\end{array}
\right)\,.
\end{eqnarray}
The density of the heat mode in the 4-mode regime is expressed in terms of hydrodynamic eigen modes as 
\begin{equation}
\epsilon-\bar{w}(\rho_R+\rho_L)=\frac{T}{2}\sqrt{\frac{\pi}{3u}}
(\Psi_1+\Psi_2+\Psi_3+\Psi_4)\,.
\end{equation}
All the coupling constants in this regime are of the same order
\begin{equation}
\lambda_{1}=
\frac{iTu^{3/2} \sqrt{3/\pi}}{4}
\left(
\begin{array}{cccc}
 5 & -1 & 1 & -1 \\
 -1 & -3 & 1 & -1 \\
 1 & 1 & 1 & 1 \\
 -1 & -1 & 1 & -3 \\
\end{array}
\right)\,,
\end{equation}
\begin{equation}
\lambda_{2}=
\frac{iTu^{3/2} \sqrt{3/\pi}}{4}
\left(
\begin{array}{cccc}
 3 & 1 & 1 & -1 \\
 1 & -5 & 1 & -1 \\
 1 & 1 & 3 & -1 \\
 -1 & -1 & -1 & -1 \\
\end{array}
\right)\,,
\end{equation}
\begin{equation}
\lambda_{3}=
\frac{iTu^{3/2} \sqrt{3/\pi}}{4}
\left(
\begin{array}{cccc}
 1 & 1 & 1 & 1 \\
 1 & -3 & -1 & -1 \\
 1 & -1 & 5 & -1 \\
 1 & -1 & -1 & -3 \\
\end{array}
\right)\,,
\end{equation}
\begin{equation}
\lambda_{4}=
\frac{iTu^{3/2} \sqrt{3/\pi}}{4}
\left(
\begin{array}{cccc}
 3 & -1 & 1 & 1 \\
 -1 & -1 & -1 & -1 \\
 1 & -1 & 3 & 1 \\
 1 & -1 & 1 & -5 \\
\end{array}
\right)\,.
\end{equation}

\subsection{3- mode regime}
In the three mode regime zero modes of the collision integral are given by
\begin{eqnarray}
{\boldsymbol \phi}^T=T^{-1} (\mu,v,-1).
\end{eqnarray}
The response function
\begin{equation}
\label{chi3}
\chi_{3} =\frac{T}{\pi u}
\left(
\begin{array}{ccc}
 \frac{49 \pi ^4 T^4}{24 m^4 u^8}+\frac{\pi ^2 T^2}{2 m^2 u^4}+1 & 0 & -\frac{7 \pi ^4 T^4}{6 m^3 u^6}-\frac{\pi ^2 T^2}{3 m u^2} \\
 0 & -\frac{7 \pi ^4 T^4}{24 m^2 u^6}+m^2 u^2-\frac{\pi ^2 T^2}{6 u^2} & 0 \\
 -\frac{\pi ^2 T^2}{3 m u^2} & 0 & \frac{\pi ^2 T^2}{3} \\
\end{array}
\right)\,.
\end{equation}
The dissipationless part of the current matrix is given by
\begin{equation}
\label{M3}
M_{3} =\frac{ikT}{\pi}
\left(
\begin{array}{ccc}
 0 & -\frac{7 \pi ^4 T^4}{24 m^3 u^7}-\frac{\pi ^2 T^2}{6 m u^3}+m u & 0 \\
 -\frac{7 \pi ^4 T^4}{24 m^3 u^7}-\frac{\pi ^2 T^2}{6 m u^3}+m u & 0 & \frac{7 \pi ^4 T^4}{30 m^2 u^5}+\frac{\pi ^2 T^2}{3 u} \\
 0 & \frac{7 \pi ^4 T^4}{30 m^2 u^5}+\frac{\pi ^2 T^2}{3 u} & 0 \\
\end{array}
\right)\,.
\end{equation}
The rotating matrix
\begin{eqnarray}
R_3\simeq
\left(
\begin{array}{ccc}
 -\frac{\pi ^{3/2} T}{\sqrt{3} m u^{3/2}} & 0 & \frac{\sqrt{\frac{3}{\pi }} \sqrt{u}}{T}-\frac{\sqrt{3} \pi ^{3/2} T}{2 m^2 u^{7/2}} \\
 \sqrt{\frac{\pi }{2}} \sqrt{u} & -\frac{\sqrt{\frac{\pi }{2}}}{m \sqrt{u}} & \frac{\sqrt{2 \pi }}{m u^{3/2}} \\
 \sqrt{\frac{\pi }{2}} \sqrt{u} & \frac{\sqrt{\frac{\pi }{2}}}{m \sqrt{u}} & \frac{\sqrt{2 \pi }}{m u^{3/2}} \\
\end{array}
\right)\,.
\end{eqnarray}
The velocities of the eigenmodes in 3-mode regime are
\begin{eqnarray}&&
u_1=0,u_2\simeq -u-\frac{\pi ^2 T^2}{3 m^2 u^3},u_3\simeq u+\frac{\pi ^2 T^2}{3 m^2 u^3}\,.
\end{eqnarray}
The heat current
\begin{equation}
\epsilon-\bar{w}\rho=\sqrt{\frac{\pi}{3u}} T \Psi_1\,.
\end{equation}
The coupling constants
\begin{equation}
\lambda_1=
\left(
\begin{array}{ccc}
 0 & \frac{3 i \sqrt{\frac{\pi }{2}} T^2}{m \sqrt{u}} & -\frac{3 i \sqrt{\frac{\pi }{2}} T^2}{m \sqrt{u}} \\
 \frac{3 i \sqrt{\frac{\pi }{2}} T^2}{m \sqrt{u}} & \frac{i \sqrt{3} \pi ^{3/2} T^3}{m^2 u^{5/2}} & 0 \\
 -\frac{3 i \sqrt{\frac{\pi }{2}} T^2}{m \sqrt{u}} & 0 & -\frac{i \sqrt{3} \pi ^{3/2} T^3}{m^2 u^{5/2}} \\
\end{array}
\right)\,,
\end{equation}
\begin{equation}
\lambda_2=\left(
\begin{array}{ccc}
 \frac{21 i \sqrt{2} \pi ^{5/2} T^4}{5 m^3 u^{9/2}} & \frac{7 i \pi ^{3/2} T^3}{2 \sqrt{3} m^2 u^{5/2}} & -\frac{7 i \pi ^{3/2} T^3}{2 \sqrt{3} m^2 u^{5/2}} \\
 \frac{7 i \pi ^{3/2} T^3}{2 \sqrt{3} m^2 u^{5/2}} & \frac{7 i \pi ^{5/2} T^4}{30 \sqrt{2} m^3 u^{9/2}} & \frac{21 i \pi ^{5/2} T^4}{10 \sqrt{2} m^3 u^{9/2}} \\
 -\frac{7 i \pi ^{3/2} T^3}{2 \sqrt{3} m^2 u^{5/2}} & \frac{21 i \pi ^{5/2} T^4}{10 \sqrt{2} m^3 u^{9/2}} & -\frac{133 i \pi ^{5/2} T^4}{30 \sqrt{2} m^3 u^{9/2}} \\
\end{array}
\right)\,,
\end{equation}
\begin{equation}
\lambda_3=\left(
\begin{array}{ccc}
 -\frac{21 i \sqrt{2} \pi ^{5/2} T^4}{5 m^3 u^{9/2}} & \frac{7 i \pi ^{3/2} T^3}{2 \sqrt{3} m^2 u^{5/2}} & -\frac{7 i \pi ^{3/2} T^3}{2 \sqrt{3} m^2 u^{5/2}} \\
 \frac{7 i \pi ^{3/2} T^3}{2 \sqrt{3} m^2 u^{5/2}} & \frac{133 i \pi ^{5/2} T^4}{30 \sqrt{2} m^3 u^{9/2}} & -\frac{21 i \pi ^{5/2} T^4}{10 \sqrt{2} m^3 u^{9/2}} \\
 -\frac{7 i \pi ^{3/2} T^3}{2 \sqrt{3} m^2 u^{5/2}} & -\frac{21 i \pi ^{5/2} T^4}{10 \sqrt{2} m^3 u^{9/2}} & -\frac{7 i \pi ^{5/2} T^4}{30 \sqrt{2} m^3 u^{9/2}} \\
\end{array}
\right)\,.
\end{equation}
Note that self coupling of the heat mode vanishes
\begin{equation}
\lambda_{1,1,1}=0\,.
\end{equation}

\section{Hierarchy of time and length scales}
In this section we evaluate the length scales that appear in the problem. 
We divide them into two groups.  The first one are 
the scales that appear on the level of Bolztmann equation, such as $\tau_F$, $L_4$ and $L_U$,
see Eqs(1), (3) and (4) in the main text.   The second scale is $L_*$ and it appear due to renormalization of the bare parameters by fluctuations, see the discussion bellow Eq(12) in the main text. 
\subsection{Scales of Boltzmann equation}
In this section we summarise different time and length scales in the problem.
The shortest time scale is a scale is determined by  three-particle collisions.
The corresponding matrix element of three fermion collision\cite{bfduality} is given by
\begin{equation}
W^{k^\prime k_2^\prime k_3^\prime}_{k k_2 k_3}= \frac{\gamma l^4}{m_{*}^2 u} 
(k_2-k)^2 (k_2^\prime- k^\prime)^2~\delta( k_2 +k -k _2^\prime -k ^\prime)~ \delta(k_3 -k_3^\prime) \nonumber.
\end{equation}
Here $\gamma=\alpha^2(1+\alpha^2)$ and $\alpha=\frac{1-K^2}{3+K^2}$ and $l$ is a radius of short range interaction.
On the level of diagonal approximation, we find the decay rate  
\begin{equation}
 \frac{1}{\tau_F(k)} =\begin{cases}
       \frac{\gamma l^4 T k^6}{m_{*}^2 u^2}, \hspace{2cm}k > \frac{T}{u},\\ \\
       \frac{\gamma l^4 T^7}{m_{*}^2 u^8}, \hspace{2.2cm}k < \frac{T}{u}.\\
   \end{cases}
   \label{fdecay}
   \end{equation}
reproducing the results of Ref. \cite{imambekov11, bfduality,Khodas2007,Lunde2007}. 
For spatial scales longer than $u\tau_F(T/u) $ electrons form a fluid with six chiral modes.
The separation between the chiral sectors is not exact and breaks and scale 
$L_4 \sim u\tau_F(p) (\frac{p}{\delta p})^2$ where $ p\sim T/u, \delta p \sim \frac{T^2}{\epsilon_F u}$ is the momentum transfer between left and right branch. Thus, left and right momentum and energies are no longer conserved hydrodynamic variable beyond the scale  
\begin{equation}
L_4 = u\tau_F(p) \left( \frac{m_* u^2}{T}\right)^2 = \frac{ m_*^4 u^{13}}{\gamma l^4 T^9}.
\end{equation}
Finally, if one accounts for merger  of the chiral branches  at the bottom of the energy, one find a time of  
the equilibration for  the number of particles between 
the left and right fermions. Within the bosonic description
this process corresponds to the Umklapp scattering, with a length  scale
\cite{matprb12,mick2010,Mat2012,mat2014}
\begin{equation}
L_U\sim u T^{-3/2}\epsilon_F^{1/2}e^{\frac{\epsilon_F}{T}}\,.
\end{equation} 

\subsection { From weak to strong coupling fixed point using MSR formalism}
\label{rgmsr}
To determine the scale $L_*$ we analyse the RG flow and estimate the value of dimensionless coupling constants.
To perform weak coupling RG we focus on the action (\ref{spsi}) and (\ref{spsiint}).
Ignoring the interaction between different modes the action in 6-mode and 4-mode regimes can be mapped onto KPZ model\cite{KPZ}.
By defining $\psi=-i\partial_xh$ and $p=\bar{\psi}/u$, one finds the standard action of KPZ model \cite{kamenev}
\begin{eqnarray}
iS=p\left(\partial_t-u\partial_x+\tilde{D}\partial_x^2\right)h-4T \tilde{D}p^2+\frac{\lambda}{T^2} p(\partial_xh)^2\,.
\end{eqnarray}
 The RG equation of this model can be written in terms of dimensionless coupling constant 
 \begin{eqnarray}
 \label{g0}
 g=\frac{\lambda^2 u\tau_F}{T^3 \tilde{D}^2}\,,
 \end{eqnarray}
 \begin{equation}
 \partial_l g = g-2 g^2\,,
 \end{equation}
 where $l=\log(L/L_0)$, and $L_0$ is an ultraviolet length cutoff.
 The solution reads 
 \begin{equation}
 g(l) =\frac{e^l g_0}{1+ 2 g_0 (e^l -1)}\,,
 \label{geffsol}
 \end{equation}
where $g_0$ is a bare value of dimensionless coupling constant.
Estimating (\ref{g0}) 
for the 4-mode regime, we find
$g_0\simeq \left(\frac{T}{T_{BF}}\right)^2 \ll 1$.
This means that the fermionic hydrodynamics is in the weak coupling regime
at spatial scales of the order of $L_4$. This implies that computations performed within the linearized
hydrodynamics in this regime are justified. 

The system enters into  strong coupling limit at $g(L_*) \sim 1$, i.e. at 
$L_*\sim L_4/g_0 \sim L_4\left(\frac{T_{BF}}{T}\right)^2 \gg L_4$. 
 \section{Asymptotics of the distribution functions}
In this section we study the asymptotic form of the pulses in different regimes, analysing the self-consistent mode-coupling equation. 
The results of this section are used in Eqs.(13), (14),(15) of the main text.
We start with the 4-mode regime.
\subsection{4-mode regime}
To compute the asymptotic, we first analyse the impact of the interaction between two modes. 
The "slow" mode  propagating with velocities $u_k$ and  a "fast" mode moving a velcity  $u_l$ in the same direction. To be concrete, let us focus on the  right (fron) tail of the mode $k$. 
 In the reference frame that moves with velocity $u_k$,  
 the tail of this mode controlled by the coupling to the mode $l$ is governed by the equation
\begin{eqnarray}
\label{eq_,ode_coupling1}
\partial_tf_k(x,t) \simeq\frac{\lambda^2_{kll}}{T^5}\int_{-\infty}^\infty dy\int_0^t ds f_k(x-y,t-s)\partial_y^2f_l^2(y,s)\,.
\end{eqnarray}
Here we omit the self-coupling and diffusion terms, which play no role far from the maximum.
Due to the scale separation, this equation can be further simplified as follows:
\begin{eqnarray}
\label{eq_,ode_coupling2}
\partial_tf_k(x,t) \simeq
\frac{\lambda^{2}_{kll}}{T^2\lambda^{2/3}_{lll}}
\partial_x^2 \int_0^t \frac{ds}{s^{2/3}} f_k(x-\Delta u_{kl}s,t-s)\,.
\end{eqnarray}
Here we used the fact that the integration over spatial coordinate is limited to a region much smaller than the separation between the peaks, yielding $\int dy f_l^2(y,s)=\frac{T^3}{\lambda_{lll}^{2/3} s^{2/3}}$.
The slow mode can be approximated by its tail estimated at the peak of the fast mode. The latter is  located at the point 
$y \sim \Delta u_{kl} s$.  
As distance $x$ is smaller that the separation between the pulses $s\Delta u_{kl}$, one may neglect $s\sim x/(\Delta u_{kl})$ compared with $t$. Therefore, Eq.(\ref{eq_,ode_coupling2}) can be further simplified, yileding
\begin{eqnarray}
\partial_tf_k(x,t) \simeq\frac{\lambda_{kll}^{2}}{\lambda_{lll}^{2/3}T^2} \partial_x^2 \int_0^t \frac{ds}{s^{2/3}} f_k(x-\Delta u_{kl}s,t)\,.
\end{eqnarray}
In the Fourier space, this  reads
\begin{equation}
\label{tails}
\partial_t f_k(q,t)=-\frac{\lambda_{kll}^{2}}{\lambda^{2/3}_{lll}T^2}q^2\int_0^t\frac{ds}{s^{2/3}}e^{-ik\Delta u_{kl}s}f_k(q,s)\,.
\end{equation}
Therefore one can look for a solution of the form
\begin{eqnarray}
f_k(q,t)=T h(q^\gamma t)\,.
\end{eqnarray}
Plugging this anzats into Eq.(\ref{tails}), one finds  
\begin{eqnarray}
h(z) \simeq \exp\left(-\frac{\lambda_{kll}^{2}z}{\lambda_{lll}^{2/3}T^2 (\Delta u_{kl})^{1/3}} \right)\,.
\end{eqnarray}
Fourier transforming back, we get the result for the tail of the mode $f_k$:
\begin{equation}
\label{tail1}
f_k(x,t)\simeq \frac{\lambda_{kll}^{2} t x^{-(\gamma+1)}}{T\lambda^{2/3}_{lll} (\Delta u_{kl})^{1/3}}\,.
\end{equation}
Substituting $\gamma=5/3$ and the values mode velocities and coupling constant, we finally find that modes propagating in same direction mutually induce  a tail that scales as
\begin{eqnarray}
\label{tail4modes}
f_k(x,t)\sim (m u^7)^{1/3} t x^{-8/3} \,,
\end{eqnarray}
where $x$ is the distance from the center of the peak.  For each of the modes, the tail is on the side directed to the other mode. 
A similar analysis yields the tail between oppositely moving modes:
\begin{eqnarray}
f_k(x,t)\sim (Tu^5)^{1/3} t x^{-8/3}.
\end{eqnarray}
\subsection{3-mode regime}
We now apply similar arguments for the 3-mode regime.
In this case the tails of the static heat mode ($j=1$) is governed by Eq.(\ref{tail1}). Substituting the coupling constant for this regime, we  find
\begin{equation}
\label{tail3}
f_1(x,t)\simeq \frac{T^{7/3}}{m^2u^{7/3}}tx^{-8/3}
\qquad  {\rm for} \ \ x \gg \left(\frac{T^4t^3}{m^6 u^7}\right)^{1/5}.
\end{equation}
The rear tails of sound modes ($j=2,3$) in this regime are formed due to interaction between sound modes, with the result
\begin{equation}
f_{2/3}(x,t) \sim \frac{T^{13/3}}{m^4u^{19/3}}tx^{-8/3}\,.
\end{equation}
\section{Thermal conductivity}
In this section  we compute thermal conductivity via Kubo formula in 6,4, and 3-modes regimes.
We establish the formal relation between the linear response thermal conductivity and the pulse propagation problem.
The results of this section are used in Eqs.(18), (19) and (20) of the main text.
\subsection{ Kubo formula for thermal conductivity}
Here we review  the Kubo formula approach for multi-component fluid,  focusing in more details on the thermal conductivity. 
The starting point is that the  fluid state assumes a local equilibrium, therefore the response function to external 
forces can be presented in the linear response formalism\cite{Kadanoff,Kovtun}. 
For multi-component fluid this refers to equilibration of the corresponding modes. 
For the brevity of notation, we suppress the chirality indexes, and restore them when needed.
In the presence of the external time-dependent perturbation $\hat{V}$ the Hamiltonian of the fluid is given by
\begin{eqnarray}
\label{6modes}
\hat{H}(t)=\hat{H}_0+\hat{V}(t)\,.
\end{eqnarray}
The perturbation can be expressed as time and space dependent thermodinamic potentials
\begin{eqnarray}
\hat{V}(t)=-\int dx \bigg\{\frac{\delta T(x,t)}{T}
[\hat{\epsilon}(x,t)-\mu\hat{\rho}(x,t)]+\delta\mu(x,t)\hat{\rho}(x,t)+v(x,t)\hat{g}(x,t)
\bigg\} \,.
\end{eqnarray}
The expectation value of a generic operator $J_i$ at a time $t$ is given as an average with respect to equilibrium density matrix:
\begin{eqnarray}
\label{linear_resp}
\langle \hat{J}_i\rangle (t)=\frac{i}{\hbar}
\int_{-\infty}^tdt'{\rm Tr} \big\{\hat{\rho}_0[\hat{V}^I(t'),\hat{J}^I_i(t)]
\big\} \,.
\end{eqnarray}
We now define the retarded current-current correlation function
\begin{eqnarray}
\label{eq_K}
K_{ij}(\omega,k)=\frac{i}{\hbar}\int_{-\infty}^{\infty} dx\int_0^\infty dt e^{-ikx+i\omega t}\langle[ \hat{J}_i(x,t),\hat{J}_j(0,0)] \rangle \,.
\end{eqnarray}
 By using (\ref{linear_resp}) one can show that linear-response coefficients $L_{ij}$ can be expressed as
\begin{eqnarray}
\label{eq_L}
L_{ij}(k)=\frac{1}{-i\omega}
\bigg[K_{ij}(\omega,k)-K_{ij}(\omega=0,k\rightarrow 0)\bigg]\,.
\end{eqnarray}
The Kubo framework can be used for computing any linear response coefficients, and in particular thermal conductivity. 
To do it, one needs to define a thermal current. In a general many-body problem it can be computed by coupling the system to the gravitational field. For the fluid, this reduces to a much simpler expression\cite{Luttinger1964} 
\begin{eqnarray}
\label{IT}
\hat{J}_T=\hat{J}_E-\bar{w} \hat{J}_\rho\,.
\end{eqnarray}
Here $J_E$ and $J_\rho$ are energy current and particle currents that are determined by energy conservation and particle conservation, $\bar{w}\simeq \pi^2T^2/4mu^2$ is enthalpy of the fluid per one electron (without the Fermi energy part).
The Kubo formula for thermal conductivity reads 
\begin{eqnarray}
\sigma_T(\omega,k)=\frac{1}{i\omega T}\int_0^L dx \int_0^{\infty}dt e^{-ikx+i\omega t}\langle [\hat{J}_T(x,t),\hat{J}_T(0,0)]\rangle\,.
\end{eqnarray}
Using the continuity equation for energy,
\begin{equation}
\frac{\partial \epsilon}{\partial t}=-{\rm div} J_E \,,
\end{equation}
and particle number,
\begin{eqnarray}
\frac{\partial \rho}{\partial t}=-{\rm div} J_\rho \,,
\end{eqnarray}
one can expressed the heat-conductivity as
\begin{equation}
\sigma_T(\omega,k)=\frac{1}{T}\frac{\omega}{k^2}\langle[\hat{\epsilon}-\bar{w}\hat{\rho},\hat{\epsilon}-\bar{w}\hat{\rho} \rangle^{\rm ret}_{\omega,k}\,.
\end{equation}
This can be cast in terms of the response function 
\begin{eqnarray}
\chi^{\rm ret}_{i,j;\eta}(x,t)=-i\theta(t)\langle [\hat{{\bf q}}_{i,\eta}(x,t), \hat{{\bf q}}_{j,\eta}(0,0)] \rangle \,.
\end{eqnarray}
Specifically, for the six-mode regime
\begin{eqnarray}
\label{sigmaKubo}
\sigma_{T}(\omega,k)=\frac{i\omega}{k^2 T^2}\sum_\eta
\bigg[\chi^{\rm ret}_{33;\eta}(\omega,k)+\bar{w}^2\chi^{\rm ret}_{11;\eta}(\omega,k)-2\bar{w}\chi^{\rm ret}_{31;\eta}(\omega,k)\bigg] \,,
\end{eqnarray}
and similarly for 4-mode and 3-mode regimes.
On the Gaussian level, the correlation function (\ref{sigmaKubo}) can be easily computed.
Indeed, in this case 
\begin{eqnarray}
\label{chi_assymp}
\chi^{\rm ret}(\omega,k\rightarrow 0)=\frac{1}{i\omega T^2}M(k).
\end{eqnarray}

\subsection{Analysis of thermal conductivity in different regimes}
In this section we employ the Kubo formula for the thermal conductivity, 
and compute it for all possible regimes. 
We start with 6-modes case.
\subsubsection{6-mode regime}
In 6-mode case Eq. (\ref{sigmaKubo}) combined with Eq.(\ref{chi_assymp}) yields
\begin{eqnarray}
\sigma_T(\omega,k)=\frac{1}{T}\bigg[L_{33}(k)+\bar{w}^2L_{11}(k)-\bar{w}L_{31}(k)-\bar{w}L_{13}(k)\bigg]\,.
\end{eqnarray}
in agreement with linear response computations done within Boltzmann equation, see Eq. (\ref{thermal_current}). 

After computing the response matrix one finds
\begin{equation}
\label{sigma_6}
\sigma_T(\omega=0,k)=
 \frac{\pi^2T}{3ik}+\frac{\pi^2u}{6l^4 m^2 T^2}\,.
\end{equation}
Here the first term correspond to the ballistic transport and yield the $\sigma_T =\frac{\pi T}{6}L$ upon a replacement 
$k \rightarrow 2\pi/L$ for the ideal quantum wire of the length $L$.
The second term to the diffusion spreading of the heat.

To discuss effects of renormalization is convenient to connect the thermal conductivity to the problem of pulse propagation.
In the 6-mode regime, the thermal conductivity can be written in terms of Keldysh correlation function $f_{ij}$ as
\begin{equation}
\sigma_T(\omega,k)=\frac{\pi^2\omega^2}{12 uk^2}
\left(\sum_{j_1,j_2=1}^2f_{j_1,j_2}(\omega,k)+\sum_{j_1,j_2=4}^5f_{j_1,j_2}(\omega,k)\right)\,.
\end{equation}
By emploing  Eq.(\ref{psi_Keldysh}) one readily reproduces the d.c. limit the real part of the conductivity Eq.(\ref{sigma_6}).
\begin{equation}
{\rm Re \sigma}(\omega\rightarrow 0,k)=\frac{\pi^2T}{6 u}
[\tilde{D}_{11}+\tilde{D}_{12}+\tilde{D}_{21}+\tilde{D}_{22}]
=\frac{\pi^2 u}{6l^4 m^2T^2}\,.
\end{equation}
Because the renormalization is weak in this regime, this is a good (up to a small correction) estimate of thermal conductivity.
\subsubsection{4-mode regime}
In this regime the bare thermal conductivity is given by  
\begin{equation}
\label{sigma_4}
\sigma_T(\omega=0,k)\sim \frac{\pi^2T}{3ik}+\frac{u^5}{l^4 T^4}.
\end{equation}
The imaginary (ballistic) part is protected by momentum conservation and therefore is uncahnged, compared with 6-mode regime. The real part of the thermal conductivity is parametrically bigger.
We now cast the Kubo formula in terms of Keldysh correlation functions, 
\begin{equation}
\label{thermal_conductivity_4modes}
\sigma(\omega,k)=\frac{\pi^2\omega^2}{12 uk^2}\sum_{j_1,j_2=1}^4f_{j_1,j_2}(\omega,k)\,.
\end{equation}
Computed on the Gaussian level, this yields
\begin{equation}
{{\rm Re}\, \sigma(\omega=0,k)}=\frac{\pi^2 T}{6u}\sum_{j_1,j_2=1}^4\tilde{D}_{j_1,j_2}\sim \frac{u^5}{l^4 T^4} \,.
\end{equation}
in agreement with Eq.(\ref{sigma_4}).

We now take into account the self-renormalization effects, by employing Eq.(\ref{thermal_conductivity_4modes})
with the pulse shape found from the self-consistent equation (8).
Both the heads and tails of pulses are modified by interaction, and contribute to the thermal conductivity.
The head of each pulse is governed by KPZ function 
$f(x,t) \simeq \frac{T^2}{(\lambda t)^{2/3}} 
f_{\rm KPZ}\left(\frac{T(x-ut)}{(\lambda t)^{2/3}}\right)$.
This yields the contribution to thermal conductivity that scales with frequency as
\begin{equation}
\label{head}
{{\rm Re}\, \sigma(\omega,k=0)}\sim \frac{\lambda^{4/3}}{Tu}\omega^{-1/3}\simeq uT^{1/3}\omega^{-1/3}\,.
\end{equation}
In addition, the tails of the distribution function(\ref{tail4modes}) contribute
 \begin{equation}
{\rm Re} \, \sigma(\omega,k)\sim (mu^4)^{1/3}k^{-1/3}\,.
\end{equation}
Substituting $k=\omega/\Delta u$ one gets
\begin{equation}
\label{tails}
{\rm Re} \, \sigma (\omega,k=0)\sim uT^{1/3}\omega^{-1/3}\,.
\end{equation}
Thus, we see that the contribution to the thermal conductivity from the head (\ref{head}) and the tails (\ref{tails}) in four mode regime have the same order.
\subsubsection{3-mode regime}
The value of the thermal conductivity in the whole 3-mode regime is strongly renormalized by interaction between the modes, 
so that its bare value has no significance. 
We thus express the conductivity in terms of pulse correlation functions
\begin{equation}
\label{heat_conductivity}
\sigma (\omega,k)=\frac{\pi^2\omega^2}{12u k^2}f_1(\omega,k)\,.
\end{equation}
Due to the lack of self interaction there  is no anomalous peak broadening of the heat mode in the three mode regime, 
and the thermal conductivity is determined solely 
by the tails of the pulse. 
Substitution of  the asymptotics Eq.(\ref{tail3}) into Eq.(\ref{heat_conductivity}) yields
\begin{equation}
{\rm Re \,\sigma} (\omega,k) \sim \frac{T^{7/3}}{u^{10/3} m^2}k^{-1/3}\,.
\end{equation}
Substituting $k=\omega/u$, we find
\begin{equation}
{\rm Re} \,\sigma (\omega,k=0) \sim \frac{T^{7/3}}{u^3m^2}\omega^{-1/3}\,.
\end{equation}

\maketitle
 

\begin{thebibliography}{99}
\bibitem{footnote1}The fermionic quasiparticles introduced  in this way are advantageous over the original electrons of the model because their interaction vanishes in the low-energy limit. In particular, they have flat density of states at the Fermi surface.
\bibitem{footnote2}
While it is difficult to give a rigorous justification to the mode-coupling treatment of the 1D fluctuation hydrodynamics, this is known to produce the results that are in excellent agreement with numerical simulations and exact solution in the case of the paradigmatic KPZ theory.
\end{thebibliography}
\end{document}